\documentclass[aps,manuscript,showpacs,psfig]{revtex4}
\usepackage{graphicx}
\usepackage{psfig}

\begin{document}

\baselineskip 0.8cm

\title{Spin transfer and polarization of antihyperons in lepton induced reactions}
\author{Dong Hui, Zhou Jian and Liang Zuo-tang}
\affiliation{Department of Physics, Shandong University,
Jinan, Shandong 250100, China}


\begin{abstract}

We study the polarization of antihyperon in 
lepton induced reactions such as 
$e^+e^-\to\bar H+X$ and $l+p\to l'+\bar H+X$ 
with polarized beams using different 
models for spin transfer in high energy fragmentation processes. 
We compare the results with the available data and those for hyperons. 
We make predictions for future experiments. 

\end{abstract}


\maketitle

\newpage

\section{Introduction}

The polarizations of hyperons 
have been widely used to study various aspects of spin effects
in high energy reactions, in particular, 
the spin-dependent fragmentation functions for their 
self spin-analyzing parity violating decay \cite{tdlee}.
One of the important aspects in this connection is the 
spin transfer in high energy fragmentation processes. 
Here, it is of particular interest to know whether 
SU(6) wave-function or the results drawn from 
polarized deeply inelastic lepton-nucleon scattering and related data 
should be used in connecting the polarization of the fragmenting 
quark and that of the produced hadrons. 
Clearly such study can provide us with useful information 
on hadronization mechanism and the spin structure of hadron.

Theoretical calculations of hyperon polarizations in different reactions 
have been carried out using different 
models \cite{GH93,BL98,Kot98,Florian98,LL00,LXL01,XLL02,LL02,
Ma:1999wp,Ma:1998pd,BQMa01,Ma:2000cg,Ma:2000uu,Ellis2002}. 
The results show, in particular, that 
it is possible to differentiate 
different pictures by measuring the polarizations of 
different hyperons in $e^+e^-$ annihilations, 
polarized deeply inelastic lepton-nucleon scattering at high energies, 
and in the high transverse momentum regions in polarized $pp$ collisions. 
They show also that the decay contributions to $\Lambda$ polarization are 
usually high and should be taken into account in the calculations, 
and that the contamination from the fragmentation of remnant of target 
in deeply inelastic scattering at relatively low energies 
such as the CERN NOMAD or DESY HERMES energies
are very important. Both of them have to be taken into account in comparing 
theoretical predictions with the experimental results.

Presently, data in this connection are already available for 
$\Lambda$ polarization in $e^+e^-$ annihilation 
at the $Z^0$ pole \cite{ALEPH96,OPAL98}, 
in deeply inelastic scattering using neutrino beam \cite{NOMAD00} 
and that using electron or muon beam \cite{HERMES,E665,COMPASS}. 
But the amount of data and the statistics of them are still 
not high enough to judge which picture is better. 
More complementary measurements are necessary and 
some of them are underway. 
In this connection, it is interesting to note that 
the NOMAD and COMPASS Collaboration at CERN has measured not only 
the polarization of $\Lambda$ but also that of $\bar\Lambda$ 
\cite{NOMAD00,COMPASS}.
Since the polarization of antihyperon in semi-inclusive 
deeply inelastic lepton-nucleon scattering is definitely more 
sensitive to the nucleon sea, it is thus instructive to make more 
detailed study in this direction not only to get more information 
on spin transfer in fragmentation but also on nucleon spin structure. 

In this paper, we make calculations of the polarizations of 
antihyperons in lepton induced reactions such as 
$e^+e^-\to Z^0\to\bar H+X$, $e^-+p\to e^-+\bar H+X$ 
and $\nu_\mu+p\to \mu^-+\bar H+X$. 
In Sec. II, 
we present the general formulas used in these calculations 
and summarize the key points of different models for 
spin transfer in fragmentation. 
We present the results for antihyperon polarizations in Sec. III and 
compare them with those for hyperons and available data.
We give a short summary and an outlook in Sec. IV.

\section{The calculation formulas}

As have been shown in \cite{LL00,LXL01,XLL02,LL02}, 
to calculate hyperon polarization in 
high energy reactions, it is necessary to take also 
the decay contribution into account. 
In this section, we present the general 
formulas for such calculations with the contribution from decay. 
We first present the formulas for a pure quark fragmentation 
in Sec. II A, then the formulas for hyperon and antihyperon 
polarization in Sec. II B. 
We summarize the key points of a few models 
for spin transfer in fragmentation processes in Sec. II C.

\subsection{Calculation formulas for a pure quark fragmentation}

We first present the formulas for a pure quark fragmentation which 
are the basis for different reactions. 
We present the formulas for $q_f\to H+X$, 
$q_f\to \bar H+X$, $\bar q_f\to H+X$ and $\bar q_f\to \bar H+X$, 
respectively, and the relations between them 
obtained from charge conjugation symmetry.  

\subsubsection{$q_f\to H+X$}

Now we consider the fragmentation process $q_f\to H+X$, 
where we use the subscript $f$ to denote the flavor of the quark, 
$H$ to denote hyperon.  
We use $D_{f}^H(z)$ to denote the fragmentation function, 
which is defined as the number density 
of $H$ produced in the fragmentation of $q_f$, 
where $z$ is the fraction of momentum of $q_f$ carried away by $H$.
We do not study the transverse momentum dependence in this paper 
so an integration over transverse momentum is understood. 
In general, if the contribution from decay is taken into account, 
$D_f^H(z)$ can be written as two parts, i.e., 
\begin{equation}
D_f^H(z)=D_f^H(z;\text{dir})+D_f^H(z;\text{dec}),
\end{equation}
where $D_f^H(z;\text{dir})$ and $D_f^H(z;\text{dec})$ denote the directly 
produced part and the decay contribution respectively.

It is clear that the decay contribution can be calculated from 
the following convolution, 
\begin{equation}
D_f^H(z;\text{dec})=\sum_j\int dz' K_{H,H_j}(z,z')D_f^{H_j}(z'),
\end{equation}
where the kernel function $K_{H,H_j}(z,z')$ is 
the probability for $H_j$ with a fractional momentum $z'$ 
to decay into an $H$ with $z$ and anything.
If we consider, as usual, only the $J^P=(1/2)^+$ octet 
and $J^P=(3/2)^+$ decuplet baryon production, 
most of the decay processes are two body decay. 
For an unpolarized two body decay $H_j\to H+M$, 
the kernel function $K_{H,H_j}(z,z')$ 
can be calculated easily. 
In this case, the magnitude of the momentum of 
the decay product in the rest frame of $H_j$ is fixed and 
it has to be isotropically distributed. 
By making a Lorentz transformation of this isotropic distribution 
to the moving frame of $H_j$, we obtain the result 
for $K_{H,H_j}(z,z')$ as given by 
\begin{equation}
K_{H,H_j}(z,\vec p_{\perp i};z',\vec p_{\perp,i}')=\frac{N}{E_j}Br(H_j\to H_iM)
\delta(p_i\cdot p_j-m_jE_i^*), 
\end{equation}
where $Br(H_j\to HM)$ is the corresponding decay branching ratio, 
$N$ is a normalization constant, 
$E^*_i$ is the energy of $H_i$ in the 
rest frame of $H_j$ which is a function of 
the masses $m_j$, $m_i$ and $m_M$ 
of $H_j$, $H_i$ and $M$.

Similarly, in polarized case, we have 
\begin{equation}
\Delta D_f^H(z)=\Delta D_f^H(z;\text{dir})+\Delta D_f^H(z;\text{dec}),
\end{equation}
where $\Delta D_f^H(z)=D_f^H(z,+)-\Delta D_f^H(z,-)$, 
and the $+$ or $-$ denotes that the produced $H$ is polarized in the 
same or opposite direction as the initial quark $q_f$. 
$\Delta D_f^H(z;\text{dir})$ and $\Delta D_f^H(z;\text{dec})$
are the corresponding quantities for directly produced $H$ 
and decay contribution. For decay contribution, 
we have \cite{ft}
\begin{equation}
\Delta D_f^H(z;\text{dec})=\sum_j\int dz' t^D_{H,H_j}
K_{H,H_j}(z,z')\Delta D_f^{H_j}(z'),
\end{equation}
where $t^D_{H,H_j}$ is the spin transfer factor for the decay 
process $H_j\to H+M$. 
$t^D_{H,H_j}$ 
is a constant which is independent of 
the process where $H_j$ is produced. It is completely 
determined by the decay process. For different decay processes, 
$t^D_{H,H_j}$ can be found, e.g., in Table II of \cite{LL00}.

The unknowns left now are $D_f^H(z;\text{dir})$ and 
$\Delta D_f^H(z;\text{dir})$. 
They are determined by the hadronization mechanism 
and the structure of hadrons. 
Presently, we can calculate $D_f^H(z;\text{dir})$ 
using a hadronization model 
or using a parametrization of fragmentation functions.
For $\Delta D_f^H(z;\text{dir})$, 
there are also different models in 
literature \cite{GH93,BL98,Kot98,Florian98,LL00,LXL01,XLL02,LL02,
Ma:1999wp,Ma:1998pd,BQMa01,Ma:2000cg,Ma:2000uu}. 
We will briefly summarize a few of them in Sec. II C.

\subsubsection{$\bar q_f\to H+X$}
For $\bar q_f\to H+X$, we have
\begin{equation}
D_{\bar f}^H(z)=D_{\bar f}^H(z;\text{dir})+D_{\bar f}^H(z;\text{dec}),
\end{equation}
\begin{equation}
D_{\bar f}^H(z;\text{dec})=\sum_j\int dz' K_{H,H_j}(z,z')D_{\bar f}^{H_j}(z'),
\end{equation}
\begin{equation}
\Delta D_{\bar f}^H(z)=\Delta D_{\bar f}^H(z;\text{dir})+\Delta D_{\bar f}^H(z;\text{dec}),
\end{equation}
\begin{equation}
\Delta D_{\bar f}^H(z;\text{dec})=\sum_j\int dz' t^D_{H,H_j}K_{H,H_j}(z,z')
                                     \Delta D_{\bar f}^{H_j}(z').
\end{equation}

\subsubsection{$q_f\to \bar H+X$}
For $q_f\to \bar H+X$, we have
\begin{equation}
D_f^{\bar H}(z)=D_f^{\bar H}(z;\text{dir})+D_f^{\bar H}(z;\text{dec}),
\end{equation}
\begin{equation}
D_f^{\bar H}(z;\text{dec})=\sum_j\int dz' K_{H,H_j}(z,z')D_f^{\bar H_j}(z'),
\end{equation}
\begin{equation}
\Delta D_f^{\bar H}(z)=\Delta D_f^{\bar H}(z;\text{dir})+\Delta D_f^{\bar H}(z;\text{dec}),
\end{equation}
\begin{equation}
\Delta D_f^{\bar H}(z;\text{dec})=\sum_j\int dz' t^D_{H,H_j}K_{H,H_j}(z,z')
                                     \Delta D_f^{\bar H_j}(z').
\end{equation}
Here we assume that the 
charge conjugation symmetry is applicable 
to the decay process so that the relations
$K_{\bar H,\bar H_j}(z,z')=K_{H,H_j}(z,z')$ and 
$t^D_{\bar H,\bar H_j}=t^D_{H,H_j}$ are valid. 
We assume also the validity of the charge conjugation symmetry 
for the fragmentation functions and obtain
further the following relations: 
\begin{equation}
D_f^{\bar H}(z;i)=D_{\bar f}^H(z;i),
\end{equation}
\begin{equation}
\Delta D_f^{\bar H}(z;i)=\Delta D_{\bar f}^H(z;i),
\end{equation}
for total, 
the direct ($i=dir$) and decay ($i=dec$) parts respectively.

\subsubsection{$\bar q_f\to \bar H+X$}

For $\bar q_f\to \bar H+X$, we use charge conjugation symmetry and obtain 
that 
\begin{equation}
D_{\bar f}^{\bar H}(z;i)=D_f^H(z;i),
\end{equation}
\begin{equation}
\Delta D_{\bar f}^{\bar H}(z;i)=\Delta D_f^H(z;i),
\end{equation}
for total,
the direct ($i=dir$) and decay ($i=dec$) parts respectively.
Hence, in the following, we need only to write out the 
formulas for $q_f\to H+X$ and $\bar q_f\to H+X$. 
Those for the other two cases are obtained from 
charge conjugation symmetry.

Independent of the models, we expect the following qualitative features 
for $D_f^H(z;\text{dir})$ and $\Delta D_f^H(z;\text{dir})$. 
Since hadrons containing the initial quark, i.e. the first rank 
hadrons in Feynman-Field type of cascade fragmentation models, 
usually carry a large fraction of momentum of the initial quark, 
we expect that, for large $z$ 
\begin{equation}
D^H_{q_f}(z;\text{dir})\gg D^H_{\bar q_f}(z;\text{dir}),
\end{equation} 
\begin{equation}
|\Delta D^H_{q_f}(z;\text{dir})|\gg |\Delta D^H_{\bar q_f}(z;\text{dir})|.
\end{equation} 
But for small $z$, they can be comparable. 
Since $P_H$ due to spin transfer is expected to be 
significant for large $z$, we should see that 
\begin{equation}
|P_H^{q_f\to HX}(z)|\gg  |P_H^{\bar q_f\to HX}(z)|.
\end{equation}
This implies that the qualitative behavior of $P_H$ in 
a given reaction is determined mainly by quark fragmentation 
and that of $P_{\bar H}$ is determined mainly by anti-quark fragmentation.

\subsection{Polarization of hyperon or antihyperon in lepton induced reactions}

To calculate the polarization of 
hyperon or antihyperon in a given reaction, 
we need to sum over the contributions from the fragmentation 
of quarks and anti-quarks of different flavors.
For example, for $A+B\to H+X$, we have
\begin{equation}
\frac{d\sigma}{dx_F}(AB\to HX)=\sum_f [
\frac{d\hat\sigma}{d\alpha}(AB\to q_fX) \otimes D_f^H(z)+
\frac{d\hat\sigma}{d\alpha}(AB\to \bar q_fX) \otimes D_{\bar f}^H(z)],
\end{equation}
\begin{equation}
\frac{d\Delta \sigma}{dx_F}(AB\to HX)=\sum_f [
\frac{d\hat\sigma}{d\alpha}(AB\to q_fX) P_{f}(\alpha)\otimes \Delta D_f^H(z)+
\frac{d\hat\sigma}{d\alpha}(AB\to \bar q_fX) P_{\bar f}(\alpha)\otimes 
\Delta D_{\bar f}^H(z)],
\end{equation}
where $x_F$ is the fractional momentum carried by the produced $H$; 
$\hat\sigma$ denotes the cross section for the production of  
$q_f$ or $\bar q_f$ in $A+B$ collisions and 
$\alpha$ denotes the kinematic variables describing cross section; 
$P_{f}(\alpha)\equiv (\Delta d\hat\sigma/d\alpha)/(d\hat\sigma/d\alpha)$ 
is the polarization of $q_f$, and similar for $P_{\bar f}$.
In general, they can be dependent on some kinematic variables 
hence we use $\otimes$ to denote convolutions.
Or equivalently, we have, 
\begin{equation}
N(x_F,H)=\sum_f [R_f(\alpha) \otimes D_f^H(z)+
R_{\bar f}(\alpha)\otimes D_{\bar f}^H(z)],
\end{equation}
\begin{equation}
\Delta N(x_F,H)=\sum_f [R_f(\alpha) P_f(\alpha) \otimes\Delta D_f^H(z)+
R_{\bar f}(\alpha)P_{\bar f}(\alpha)\otimes\Delta D_{\bar f}^H(z)],
\end{equation}
\begin{equation}
N(x_F,\bar H)=\sum_f [R_f(\alpha) \otimes D_{\bar f}^H(z)+
R_{\bar f}(\alpha)\otimes D_f^H(z)],
\end{equation}
\begin{equation}
\Delta N(x_F,\bar H)=\sum_f R_f(\alpha) P_f(\alpha) \otimes\Delta D_{\bar f}^H(z)+
   R_{\bar f}(\alpha)P_{\bar f}(\alpha)\otimes\Delta D_f^H(z)],
\end{equation}
where we use $N(x_F,H)$ and $N(x_F,\bar H)$ to denote 
the number density of $H$ and that of $\bar H$ 
at a given momentum fraction $x_F$ produced in the reaction 
respectively, 
$\Delta N(x_F,H)$ and $\Delta N(x_F,\bar H)$ 
to denote the corresponding differences in the polarized case.
$P_f$ and $P_{\bar f}$ are respectively the polarization of $q_f$ and 
that of $\bar q_f$; 
$R_f$ and $R_{\bar f}$ are the fractional contributions 
of $q_f\to h+X$ and $\bar q_f\to h+X$ to the whole hadronic events.
They are related to $d\hat\sigma$ by
\begin{equation}
	R_f(\alpha)=\frac{1}{\sigma_{\text{inel}}}\frac{d\hat\sigma}{d\alpha}(AB\to q_fX),
\end{equation}
where $\sigma_{\text{inel}}$ is the total inelastic cross section for hadron production 
in $A+B$ collisions. 
The final result for the polarization is given by,
\begin{equation}
P_H(x_F)=\frac{\Delta N(x_F,H)}{N(x_F,H)}.
\end{equation}
\begin{equation}
P_{\bar H}(x_F)=\frac{\Delta N(x_F,\bar H)}{N(x_F,\bar H)}.
\end{equation}

For different reactions,  
the fragmentation functions $D$'s and $\Delta D$'s
are assumed to be universal,
but the results of $R_f$, $R_{\bar f}$, $P_f$ 
and $P_{\bar f}$ are different. 
These differences lead to  
different results for the polarization of hyperons 
and/or antihyperons.
In the following, we summarize the formulas in 
different lepton induced reactions, respectively, 
and discuss the qualitative features of these results. 
The numerical results are given in Sec. III.

\subsubsection{$e^+e^-\to H({\rm or\ } \bar H)+X$}

At high energies, due to the contribution through 
weak-interaction at the $e^+e^-$ annihilation vertex 
such as $e^+e^-\to Z^0\to q_f\bar q_f$, 
the initial $q_f$ and $\bar q_f$ are longitudinally polarized. 
The magnitude of the polarization of $q_f$ and that of $\bar q_f$ 
are the same but the sign are different. 
They are constants at a given 
center of mass (c.m.) energy of $e^+e^-$ system.
Also the relative weights for the contributions of different flavors
are constants when averaging over the different jet (or initial quark) directions.  
They are the same for quarks and for anti-quarks.
Namely, we have 
\begin{equation}
P^{e^+e^-}_f=-P^{e^+e^-}_{\bar f},
\end{equation}
\begin{equation}
R^{e^+e^-}_f=R^{e^+e^-}_{\bar f}.
\end{equation}
and $R^{e^+e^-}_f$ is given by
\begin{equation}
R^{e^+e^-}_f=\frac{\sigma(e^+e^-\to q_f\bar q_f)}
                  {\sum_f \sigma(e^+e^-\to q_f\bar q_f)}.
\end{equation}
For example, for reactions at the $Z^0$-pole, 
we neglect the contribution from the annihilation via virtual photon. 
We have $P_f=-0.67$ for $f=u$ or $c$ and $P_f=-0.94$ for $f=d,s$, or $b$.
Hence, we have
\begin{equation}
N^{e^+e^-}(z,H)=\sum_f R^{e^+e^-}_f [D_f^H(z)+D_{\bar f}^H(z)],
\end{equation}
\begin{equation}
\Delta N^{e^+e^-}(z,H)=\sum_f R^{e^+e^-}_f P_f^{e^+e^-}
[\Delta D_f^H(z)-\Delta D_{\bar f}^H(z)].
\end{equation}
For antihyperons, we have
\begin{equation}
N^{e^+e^-}(z,\bar H)=\sum_f R^{e^+e^-}_f [D_f^H(z)+D_{\bar f}^H(z)],
\end{equation}
\begin{equation}
\Delta N^{e^+e^-}(z,\bar H)=\sum_f R^{e^+e^-}_f P_f^{e^+e^-}
[\Delta D_{\bar f}^H(z)-\Delta D_f^H(z)].
\end{equation}
We see that,
\begin{equation}
N^{e^+e^-}(z,\bar H)=
N^{e^+e^-}(z,H),
\end{equation}
\begin{equation}
\Delta N^{e^+e^-}(z,\bar H)=
-\Delta N^{e^+e^-}(z,H).
\end{equation}
This means that, under the charge conjugation symmetries 
for the fragmentation functions and decay processes, we have
\begin{equation}
P^{e^+e^-}_{\bar H}(z)=-P^{e^+e^-}_H(z).
\end{equation}
This result is true independent of model for fragmentation functions. 
It is a direct consequence of the charge conjugation symmetries for 
fragmentation and decay processes.
Since we do not have the contaminations from the initial state such as 
the structure of nucleon in this reaction, this is an ideal place to test 
such charge conjugation symmetries. 

\subsubsection{$e^-+N\to e^-+H({\rm or}\ \bar H)+X$}

At sufficiently large energies and momentum transfer $Q^2$, 
hadron produced in the current fragmentation region 
of semi-inclusive deeply inelastic 
lepton-nucleon scattering such as $e^-+N\to e^-+H({\rm or}\ \bar H)+X$ 
can be considered as a pure product of fragmentation of the 
struck quark. 
In this case, the cross sections are given by
\begin{equation}
\frac{d^3\sigma}{dxdydz}(eN\to eHX)=
\frac{d\hat\sigma_{\text{Mott}}}{dy}
\sum_f e_f^2 [q_f(x,Q^2)D_f^H(z)+\bar q_f(x,Q^2)D_{\bar f}^H(z)],
\end{equation}
\begin{equation}
\frac{\Delta d^3\sigma}{dxdydz}(eN\to eHX)=
\frac{d\hat\sigma_{\text{Mott}}}{dy}
\sum_f e_f^2 [P_f^{eN}(x,y) q_f(x,Q^2)\Delta D_f^H(z)+ 
P_{\bar f}^{eN}(x,y)\bar q_f(x,Q^2)\Delta D_{\bar f}^H(z)],
\end{equation}
where $\sigma_{\text{Mott}}$ is the Mott cross section; 
$x$ is the usual Bjorken $x$; 
$y$ is the fractional energy transfer 
in the rest frame of the nucleon; 
and $z$ is the fraction of 
momentum of struck $q$ carried by $H$.
The polarization of 
the struck quark $q_f$ and that of anti-quark $\bar q_f$ 
are given by  
\begin{equation}
P^{eN}_f(x,y)=\frac{P^{(e)}_L D_L(y) q_f(x)+P^{(N)}_L \Delta q_f(x)}
{q_f(x) +P^{(e)}_L D_L(y) P^{(N)}_L \Delta q_f(x)},
\end{equation}
\begin{equation}
P^{eN}_{\bar f}(x,y)=\frac{P^{(e)}_L D_L(y) \bar q_f(x)+P^{(N)}_L \Delta \bar q_f(x)}
{\bar q_f(x) +P^{(e)}_L D_L(y) P^{(N)}_L \Delta \bar q_f(x)},
\end{equation}
for longitudinally polarized case,  
where $P_L^{(e)}$ and $P_L^{(N)}$ are polarizations of 
the incident electron and nucleon, respectively;  
$D_L(y)$ is the longitudinal spin transfer factor in $eq\to eq$. 
It is given by
\begin{equation}
D_L(y)=\frac{1-(1-y)^2}{1+(1-y)^2}.
\label{sub1eq4}
\end{equation}
In the transversely polarized case, we have
\begin{equation}
P_{fT}^{eN}(x,y)=
P^{(N)}_T \frac{\delta q_f(x)}{q_f(x)}D_T(y).
\end{equation}
\begin{equation}
P_{\bar fT}^{eN}(x,y)=
P^{(N)}_T \frac{\delta \bar q_f(x)}{\bar q_f(x)}D_T(y).
\end{equation}
where the transversal spin transfer factor $D_T(y)$ in $eq\to eq$ is given by
\begin{equation}
D_T(y)=\frac{2(1-y)}{1+(1-y)^2}.
\label{sub1eq5}
\end{equation}
We see that the relative contributions of different flavors are given by
\begin{equation}
	R^{eN}_f(x,y)=\frac{1}{\sigma_{\text{inel}}}
\frac{d\hat\sigma_{\text{Mott}}}{dy}e^2_f q_f(x,Q^2),
\end{equation}
\begin{equation}
	R^{eN}_{\bar f}(x,y)=\frac{1}{\sigma_{\text{inel}}}
\frac{d\hat\sigma_{\text{Mott}}}{dy}e^2_f \bar q_f(x,Q^2).
\end{equation}
From Eqs.(42)--(49), 
we see that both $P_f$ and $R_f$ depend on $x$ and $y$. 
We see also that, in general, 
$P^{eN}_f(x,y)\not=P^{eN}_{\bar f}(x,y)$,
$R^{eN}_f(x,y)\not=R^{eN}_{\bar f}(x,y)$,
they are equal only if 
$q_f(x,Q^2)=\bar q_f(x,Q^2)$ and 
$\Delta q_f(x,Q^2)=\Delta \bar q_f(x,Q^2)$.

For the corresponding number densities of 
$H$ or $\bar H$ in $e+N\to e+H({\rm or}\ \bar H)+X$, we have 
\begin{equation}
N^{eN}(z,H)=\sum_f \int dxdy \Bigl[
R^{eN}_f(x,y) D_f^H(z)+R^{eN}_{\bar f}(x,y)D_{\bar f}^H(z)\Bigr],
\end{equation}
\begin{equation}
N^{eN}(z,\bar H)=\sum_f \int dxdy 
\Bigl[R^{eN}_f(x,y) D_{\bar f}^H(z)+ 
      R^{eN}_{\bar f}(x,y)D_f^H(z)\Bigr],
\end{equation}
\begin{equation}
\Delta N^{eN}(z,H)=\sum_f \int dxdy 
  \Bigl[R^{eN}_f(x,y) P^{eN}_f(x,y)\Delta D_f^H(z)+ 
        R^{eN}_{\bar f}(x,y) P^{eN}_{\bar f}(x,y) \Delta D_{\bar f}^H(z)\Bigr],
\end{equation}
\begin{equation}
\Delta N^{eN}(z,\bar H)=\sum_f \int dxdy 
    \Bigl[R^{eN}_f(x,y) P^{eN}_f(x,y) \Delta D_{\bar f}^H(z)+ 
          R^{eN}_{\bar f}(x,y) P^{eN}_{\bar f}(x,y) \Delta D_f^H(z)\Bigr],
\end{equation}

Now, we compare the results for $\bar H$ with those for $H$, 
and we see the following qualitative features 
in the two different cases.

In the first case, we consider reactions at very high energies 
so that small $x$ contribution dominates. 
In this case, we can neglect the valence-quark contributions 
with good accuracy, i.e., 
$q_f(x,Q^2)\approx q_{f,s}(x,Q^2)$ and 
$\Delta q_f(x,Q^2)\approx \Delta q_{f,s}(x,Q^2)$, 
where the subscript $s$ denote ``sea.'' 
If we assume the charge conjugation symmetry in 
nucleon sea, i.e., 
$q_{f,s}(x,Q^2)=\bar q_{f,s}(x,Q^2)=\bar q_f(x,Q^2)$, and
$\Delta q_{f,s}(x,Q^2)= \Delta \bar q_{f,s}(x,Q^2)=\Delta \bar q_f(x,Q^2)$, 
we expect only a very small 
difference between quark and anti-quark distributions, i.e., 
$q_f(x,Q^2)\approx \bar q_f(x,Q^2)$, and
$\Delta q_f(x,Q^2)\approx \Delta \bar q_f(x,Q^2)$. 
Hence, we expect that, 
$P_f^{eN}(x,y)\approx P_{\bar f}^{eN}(x,y)$,
$R_f^{eN}(x,y)\approx R_{\bar f}^{eN}(x,y)$,
and finally $P_H(z)\approx P_{\bar H}(z)$. 

In the second case, we consider reactions where 
large $x$ contributions dominate. 
In this case, valence-quark contribution plays an important role
and there should be significant differences 
between quark and anti-quark distributions 
| thus significant differences between $P_H$ and $P_{\bar H}$.
To get some feeling of these differences, 
we make the following rough estimations. 

We denote 
$R^{eN}_f(x,y)=R^{eN}_{f,v}(x,y)+R^{eN}_{f,s}(x,y)$,
(where the $v$ and $s$ in the subscripts denote valence or sea contribution),
and expect the following approximate relations, 
if we neglect the flavor-dependence in the 
quark distribution functions: 
\begin{equation}
R^{ep}_{u,v}(x,y)\approx 8 R^{ep}_{d,v}(x,y),
\end{equation}
\begin{equation}
R^{eN}_{u,s}(x,y)\approx 4 R^{eN}_{d,s}(x,y)\approx 4 R^{eN}_{s,s}(x,y),
\end{equation}
\begin{equation}
R^{eN}_{\bar u,s}(x,y)\approx 4 R^{eN}_{\bar d,s}(x,y)\approx 4 R^{eN}_{\bar s,s}(x,y).
\end{equation}

These approximate relations can be used to give us some 
guidances of the qualitative features of hyperon and 
antihyperon polarizations in the reaction. 
For example, for $e^-+p\to e^-+\Lambda +X$, we note further that,
$D_u^\Lambda(z)=D_d^\Lambda(z)$ and for large $z$ 
both of them have the strangeness suppressions 
factor $\lambda \approx 0.3$ relative to $D_s^\Lambda(z)$ 
and that both $u$ and $d$ have, if any, small and 
negative contributions to 
the spin of $\Lambda$ but $s$ has large and positive contribution. 
Hence, for reactions where the valence-quark contributions 
dominate for hyperon production,  we expect 
$u\to\Lambda+X$ dominates in large $z$ region,
and $P_\Lambda$ should be small. 
But for $\bar\Lambda$, there is no contribution 
from valence-quark to the first rank particles; 
we expect that, for large $z$, 
$\bar u\to \bar\Lambda+X$ and    
$\bar s\to \bar\Lambda+X$ give comparable 
contributions to $\bar \Lambda$ ptroduction 
and $P_{\bar\Lambda}$ should be mainly determined by 
$\bar s\to \bar\Lambda+X$. 
This implies that, in this case, 
$P_{\bar\Lambda}$ should be positive and 
the magnitude is much larger than $P_\Lambda$ at large $z$ values.

We can see that the qualitative features are independent 
of the models for spin transfer 
but depend strongly on the kinematic region.

\subsubsection{$\nu_\mu+N\to\mu^-+H({\rm or}\ \bar H)+X$}

For charged current neutrino reaction, we have 
\begin{eqnarray}\nonumber
\frac{d^3\sigma}{dxdydz}(\nu_\mu N\to\mu^-HX)
&=&\sum_{f,f'} 
\frac{d\hat\sigma_0}{dy}(\nu_\mu q_f\to\mu^-q_{f'})
q_f(x,Q^2)D_{f'}^H(z)\\ \nonumber 
&+& \sum_{f,f'} 
\frac{d\hat\sigma_0}{dy}(\nu_\mu\bar q_f\to\mu^-\bar q_{f'})
\bar q_f(x,Q^2)D_{\bar f'}^H(z).
\end{eqnarray}
We do not consider top production. 
In this case we have two 
Cabbibo favored elementary processes for quarks. 
The differential cross sections are given by 
\begin{equation}
\frac{d\hat\sigma_0}{dy}(\nu_\mu d\to\mu^-u)=
\frac{d\hat\sigma_0}{dy}(\nu_\mu s\to\mu^-c)=
\frac{G^2xs}{\pi}\cos^2\theta_c,
\end{equation}
and two Cabbibo suppressed elementary processes with cross sections,
\begin{equation}
\frac{d\hat\sigma_0}{dy}(\nu_\mu d\to\mu^-c)=
\frac{d\hat\sigma_0}{dy}(\nu_\mu s\to\mu^-u)=
\frac{G^2xs}{\pi}\sin^2\theta_c,
\end{equation}
where $\theta_c$ is the Cabbibo angle.
Similarly for anti-quarks, 
we have two Cabbibo favored elementary processes,
\begin{equation}
\frac{d\hat\sigma_0}{dy}(\nu_\mu \bar u\to\mu^-\bar d)=
\frac{d\hat\sigma_0}{dy}(\nu_\mu \bar c\to\mu^-\bar s)=
\frac{G^2xs}{\pi}(1-y)^2\cos^2\theta_c,
\end{equation}
and two Cabbibo suppressed processes, 
\begin{equation}
\frac{d\hat\sigma_0}{dy}(\nu_\mu \bar u\to\mu^-\bar s)=
\frac{d\hat\sigma_0}{dy}(\nu_\mu \bar c\to\mu^-\bar d)=
\frac{G^2xs}{\pi}(1-y)^2\sin^2\theta_c,
\end{equation}
This means that 
\begin{eqnarray}\nonumber
\frac{d^3\sigma}{dxdydz}(\nu_\mu N\to\mu^-HX)
&=&\frac{G^2xs}{\pi}\Bigl\{[d(x,Q^2)\cos^2\theta_c+s(x,Q^2)\sin^2\theta_c]D_u^H(z) \\ \nonumber
             &+& [d(x,Q^2)\sin^2\theta_c+s(x,Q^2)\cos^2\theta_c]D_c^H(z) \\ \nonumber
&+&(1-y)^2[\bar u(x,Q^2)\cos^2\theta_c+\bar c(x,Q^2)\sin^2\theta_c]D_{\bar d}^H(z) \\ 
&+&(1-y)^2[\bar u(x,Q^2)\sin^2\theta_c+\bar c(x,Q^2)\cos^2\theta_c]D_{\bar s}^H(z)]\Bigr\},
\end{eqnarray}
We note further that the quarks in the final state of 
$\nu_\mu q\to \mu^-q'$ are longitudinally polarized 
with polarization $P_{q'}=-1$, 
while the anti-quarks in the final state of
$\nu_\mu \bar q\to \mu^-\bar q'$ are longitudinally polarized 
with $P_{\bar q'}=1$.
We obtain that 
\begin{eqnarray}\nonumber
\frac{d^3\Delta\sigma}{dxdydz}(\nu_\mu N\to\mu^-HX)
&=&\frac{G^2xs}{\pi}\Bigl\{-[d(x,Q^2)\cos^2\theta_c+s(x,Q^2)\sin^2\theta_c]\Delta D_u^H(z) \\ \nonumber
&-&[d(x,Q^2)\sin^2\theta_c+s(x,Q^2)\cos^2\theta_c]\Delta D_c^H(z) \\ \nonumber
&+&(1-y)^2[\bar u(x,Q^2)\cos^2\theta_c+\bar c(x,Q^2)\sin^2\theta_c]\Delta D_{\bar d}^H(z) \\ 
&+&(1-y)^2[\bar u(x,Q^2)\sin^2\theta_c+\bar c(x,Q^2)\cos^2\theta_c]\Delta D_{\bar s}^H(z)\Bigr\},
\end{eqnarray}
For antihyperon, we have
\begin{eqnarray}\nonumber
\frac{d^3\sigma}{dxdydz}(\nu_\mu N\to\mu^-\bar HX)
&=&\frac{G^2xs}{\pi}\Bigl\{[d(x,Q^2)\cos^2\theta_c+s(x,Q^2)\sin^2\theta_c]D_{\bar u}^H(z) \\ \nonumber
&+& [d(x,Q^2)\sin^2\theta_c+s(x,Q^2)\cos^2\theta_c]D_{\bar c}^H(z) \\ \nonumber
&+&(1-y)^2[\bar u(x,Q^2)\cos^2\theta_c+\bar c(x,Q^2)\sin^2\theta_c]D_d^H(z)+ \\ 
&+&(1-y)^2[\bar u(x,Q^2)\sin^2\theta_c+\bar c(x,Q^2)\cos^2\theta_c]D_s^H(z)]\Bigr\},
\end{eqnarray}
\begin{eqnarray}\nonumber
\frac{d^3\Delta\sigma}{dxdydz}(\nu_\mu N\to\mu^-\bar HX)
&=&\frac{G^2xs}{\pi}\Bigl\{-[d(x,Q^2)\cos^2\theta_c+s(x,Q^2)\sin^2\theta_c]\Delta D_{\bar u}^H(z) \\ \nonumber
&-&[d(x,Q^2)\sin^2\theta_c+s(x,Q^2)\cos^2\theta_c]\Delta D_{\bar c}^H(z) \\ \nonumber
&+&(1-y)^2[\bar u(x,Q^2)\cos^2\theta_c+\bar c(x,Q^2)\sin^2\theta_c]\Delta D_d^H(z) \\ 
&+&(1-y)^2[\bar u(x,Q^2)\sin^2\theta_c+\bar c(x,Q^2)\cos^2\theta_c]\Delta D_s^H(z)]\Bigr\}.
\end{eqnarray}

From these equations, we expect that, 
unlike that in $e^-N$ scattering, 
there should be a significant 
difference for the polarization of hyperon and that for the 
corresponding antihyperon. 
This is because, in $\nu_\mu+N\to\mu^-+H({\rm or}\ \bar H)+X$, 
(i) hyperons in the current fragmentation region are mainly 
from the fragmentation of $u$ and $c$ quarks but 
the antihyperons are from $\bar{d}$ and $\bar{s}$; and 
(ii) the helicities of the struck quarks are 
opposite to those of the anti-quarks.

To see the qualitative features more explicitly, 
we now make a qualitative analysis by taking 
the following approximations.
We keep only Cabbibo favored processes and 
neglect $\Delta D_{\bar f}^H(z)$ compared to $\Delta D_f^H(z)$.
Under these approximations, we have 
\begin{equation}
\frac{d^3\Delta\sigma}{dxdydz}(\nu_\mu N\to\mu^-HX)
\approx -\frac{G^2xs}{\pi}\cos^2\theta_c
[d(x,Q^2)\Delta D_u^H(z)+s(x,Q^2)\Delta D_c^H(z)],
\end{equation}
\begin{equation}
\frac{d^3\Delta\sigma}{dxdydz}(\nu_\mu N\to\mu^-\bar HX)
\approx \frac{G^2xs}{\pi}(1-y)^2 \cos^2\theta_c
[\bar u(x,Q^2)\Delta D_d^H(z)+\bar c(x,Q^2)\Delta D_s^H(z)].
\end{equation}
We see that the polarization of hyperons in this reaction 
is mainly determined by $\Delta D_u^H(z)$ and $\Delta D_c^H(z)$ 
while those for antihyperons are determined by  
$\Delta D_d^H(z)$ and $\Delta D_s^H(z)$. 
We thus expect the following qualitative features:

(1) For $\Lambda$, we note $\Delta D_d^\Lambda(z)=\Delta D_u^\Lambda(z)<0$ 
and the magnitude is very small, while $\Delta D_s^\Lambda(z)>0$ and the 
magnitude is large. 
There is a significant contribution from $\Lambda_c$ decay to $\Lambda$, 
but the decay spin transfer in $\Lambda_c\to\Lambda X$ is unclear. 
Hence it is very difficult to make any estimate on $P_\Lambda$.
But for $\bar\Lambda$, we expect that the qualitative feature of 
$P_{\bar\Lambda}$  in $\nu_\mu+N\to\mu^-+\bar\Lambda+X$
is mainly determined by $\Delta D_s^\Lambda(z)$.
The results should be positive and the magnitude is large for large $z$.

(2) For $\Sigma$ production,
we recall that,
from isospin symmetry, 
$\Delta D_u^{\Sigma^+}=\Delta D_d^{\Sigma^-}$ is positive and large,
while $\Delta D_d^{\Sigma^+}=\Delta D_u^{\Sigma^-}$ is very small.
The spin transfer from charmed baryon decay to $\Sigma$ 
is unknown but such decay contribution is 
relatively small compared to $\Lambda$.
We thus expect that $P_{\Sigma^+}$ is negative and the 
magnitude is large for large $z$. 
But the magnitude of $P_{\Sigma^-}$ is very small. 

We note further that 
$\Delta D_s^{\Sigma^{\pm}}$ is negative 
and the magnitude is smaller than 
$\Delta D_d^{\Sigma^-}$ [half of it in 
SU(6) and similar in DIS picture].
But there is a strange suppression for $D_d^\Sigma$ 
compared to $D_s^\Sigma$. 
We expected the contributions from 
the two terms in Eq.(66) to $P_{\bar \Sigma^+}$ are opposite 
in sign with similar magnitudes. 
These two contributions partly cancel each other 
and the final results for $P_{\bar \Sigma^+}$ 
should be small in magnitude and very sensitive to 
the quark distribution functions. 
For $P_{\bar \Sigma^-}$, the contribution from 
$\Delta D_{\bar d}^{\bar \Sigma^-}=\Delta D_d^{\Sigma^+}$ 
is very small and the results are mainly determined 
by $\Delta D_s^{\Sigma^+}$, which is negative 
and larger at large $z$. 

(3) For $\Xi$ production, charmed baryon decay contribution 
can be neglected. 
$P_{\Xi}$ is mainly determined by $-\Delta D_u^\Xi$.
Since $\Delta D_u^{\Xi^0}(z,\text{dir})$ is negative 
but the decay contribution from $\Xi^{*0}$, i.e. 
$\Delta D_u^{\Xi^0}(z,\text{dec})$, is positive, and  
the magnitude of polarization of the latter is larger, 
the final results for $P_{\Xi^0}$ can be quite sensitive to the 
hadronization model.  
But in gerenal, we expect both the magnitude of $P_{\Xi^0}$ 
and that of $P_{\Xi^-}$ to be small.

For $\bar\Xi^0$, the contribution from the first term 
of Eq.(66), i.e. that from $\Delta D_d^{\Xi^0}$, 
is much smaller than that from the second term 
since $\Xi^0$ does not contain $d$ as a valence-quark. 
For $\bar\Xi^+$, there is a contribution from the first term 
of Eq.(66), i.e. that from $\Delta D_d^{\Xi^-}$, 
but the magnitude is smaller than $\Delta D_s^{\Xi^-}$ 
[half of it in SU(6)]. 
Furthermore, because of the strangeness suppression, 
$D_d^{\Xi-}$ also is suppressed compared to $D_s^{\Xi^-}$.
We thus expect that $|\Delta D_d^{\Xi^-}|\ll \Delta D_s^{\Xi^-}$ too.
Hence for both $\bar\Xi^0$ and $\bar\Xi^+$, we expect that  
$P_{\bar\Xi}$ is mainly determined by $\Delta D_s^{\Xi}$ 
and results should be positive and large for large $z$.

These qualitative features are independent of models for 
spin transfer and can be checked by experiments.

\subsubsection{$\bar\nu_\mu+N\to \mu^++H({\rm or}\ \bar H)+X$}

For charged current reactions with 
anti-neutrino beam such as 
$\bar\nu_\mu+N\to \mu^++H({\rm or}\ \bar H)+X$, 
the contributing elementary processes are the following.
We have two 
Cabbibo favored elementary processes for quarks  
with the differential cross sections as given by 
\begin{equation}
\frac{d\hat\sigma_0}{dy}(\bar \nu_\mu u\to\mu^+d)=
\frac{d\hat\sigma_0}{dy}(\bar \nu_\mu c\to\mu^+s)=
\frac{G^2xs}{\pi}(1-y)^2\cos^2\theta_c,
\end{equation}
and two Cabbibo suppressed elementary processes with cross sections,
\begin{equation}
\frac{d\hat\sigma_0}{dy}(\bar \nu_\mu u\to\mu^+s)=
\frac{d\hat\sigma_0}{dy}(\bar \nu_\mu c\to\mu^+d)=
\frac{G^2xs}{\pi}(1-y)^2\sin^2\theta_c.
\end{equation}
Similarly for anti-quarks, 
we have two Cabbibo favored elementary processes,
\begin{equation}
\frac{d\hat\sigma_0}{dy}(\bar \nu_\mu \bar d\to\mu^+\bar u)=
\frac{d\hat\sigma_0}{dy}(\bar \nu_\mu \bar s\to\mu^+\bar c)=
\frac{G^2xs}{\pi}\cos^2\theta_c,
\end{equation}
and two Cabbibo suppressed processes, 
\begin{equation}
\frac{d\hat\sigma_0}{dy}(\bar\nu_\mu \bar d\to\mu^+\bar c)=
\frac{d\hat\sigma_0}{dy}(\bar\nu_\mu \bar s\to\mu^+\bar u)=
\frac{G^2xs}{\pi}\sin^2\theta_c.
\end{equation}
Hence, we obtain that 
\begin{eqnarray}\nonumber
\frac{d^3\sigma}{dxdydz}(\bar\nu_\mu N\to\mu^+HX)
&=&\frac{G^2xs}{\pi}\Bigl\{
   [\bar d(x,Q^2)\cos^2\theta_c+\bar s(x,Q^2)\sin^2\theta_c]D_{\bar u}^H(z) \\ \nonumber
&+&[\bar d(x,Q^2)\sin^2\theta_c+\bar s(x,Q^2)\cos^2\theta_c]D_{\bar c}^H(z) \\ \nonumber
&+&(1-y)^2[u(x,Q^2)\cos^2\theta_c+c(x,Q^2)\sin^2\theta_c]D_d^H(z) \\ 
&+&(1-y)^2[u(x,Q^2)\sin^2\theta_c+c(x,Q^2)\cos^2\theta_c]D_s^H(z)\Bigr\}, 
\end{eqnarray}
\begin{eqnarray}\nonumber
\frac{d^3\sigma}{dxdydz}(\bar\nu_\mu N\to\mu^+\bar HX)
&=&\frac{G^2xs}{\pi}\Bigl\{
   [\bar d(x,Q^2)\cos^2\theta_c+\bar s(x,Q^2)\sin^2\theta_c]D_u^H(z) \\ \nonumber
&+&[\bar d(x,Q^2)\sin^2\theta_c+\bar s(x,Q^2)\cos^2\theta_c]D_c^H(z) \\ \nonumber 
&+&(1-y)^2[u(x,Q^2)\cos^2\theta_c+c(x,Q^2)\sin^2\theta_c]D_{\bar d}^H(z) \\ 
&+&(1-y)^2[u(x,Q^2)\sin^2\theta_c+c(x,Q^2)\cos^2\theta_c]D_{\bar s}^H(z)]\Bigr\},
\end{eqnarray}
We note further that the quarks in the final state of 
$\bar\nu_\mu q\to \mu^+q'$ are longitudinally polarized 
with polarization $P_{q'}=-1$, 
while the anti-quarks in the final state of
$\bar\nu_\mu\bar q\to \mu^+\bar q'$ are longitudinally polarized 
with $P_{\bar q'}=1$.
We obtain that,
\begin{eqnarray}\nonumber
\frac{d^3\Delta\sigma}{dxdydz}(\bar\nu_\mu N\to\mu^+HX)
&=&\frac{G^2xs}{\pi}\Bigl\{
   [\bar d(x,Q^2)\cos^2\theta_c+\bar s(x,Q^2)\sin^2\theta_c]\Delta D_{\bar u}^H(z) \\ \nonumber
&+&[\bar d(x,Q^2)\sin^2\theta_c+\bar s(x,Q^2)\cos^2\theta_c]\Delta D_{\bar c}^H(z) \\ \nonumber
&-&(1-y)^2[u(x,Q^2)\cos^2\theta_c+c(x,Q^2)\sin^2\theta_c]\Delta D_d^H(z) \\ 
&-&(1-y)^2[u(x,Q^2)\sin^2\theta_c+c(x,Q^2)\cos^2\theta_c]\Delta D_s^H(z)\Bigr\}, 
\end{eqnarray}
\begin{eqnarray}\nonumber
\frac{d^3\Delta\sigma}{dxdydz}(\bar\nu_\mu N\to\mu^+\bar HX)
&=&\frac{G^2xs}{\pi}\Bigl\{
   [\bar d(x,Q^2)\cos^2\theta_c+\bar s(x,Q^2)\sin^2\theta_c]\Delta D_u^H(z) \\ \nonumber
&+&[\bar d(x,Q^2)\sin^2\theta_c+\bar s(x,Q^2)\cos^2\theta_c]\Delta D_c^H(z) \\ \nonumber 
&-&(1-y)^2[u(x,Q^2)\cos^2\theta_c+c(x,Q^2)\sin^2\theta_c]\Delta D_{\bar d}^H(z) \\ 
&-&(1-y)^2[u(x,Q^2)\sin^2\theta_c+c(x,Q^2)\cos^2\theta_c]\Delta D_{\bar s}^H(z)]\Bigr\},
\end{eqnarray}

We compare these results with those for 
$\nu_\mu+N\to\mu^-+H({\rm or}\ \bar H)+X$ presented in last subsection. 
We see that the results for 
$\bar\nu_\mu+N\to\mu^++H({\rm or}\ \bar H)+X$ are the same as  
the corresponding results $\nu_\mu+N\to\mu^-+\bar H({\rm or}\ H)+X$ 
under the exchange $q_f(x,Q^2)\leftrightarrow\bar q_f(x,Q^2)$. 
Hence, if we consider the reactions at very high energies 
where small $x$ contribution dominates so that 
$q_f(x,Q^2)\approx\bar q_f(x,Q^2)$, 
we have 
\begin{equation}
P^{\nu N}_H(z)\approx  P^{\bar\nu N}_{\bar H}(z),
\end{equation}
\begin{equation}
P^{\nu N}_{\bar H}(z)\approx  P^{\bar\nu N}_{H}(z).
\end{equation}

As we have emphasized before, 
to test different models for spin transfer in fragmentation, 
it is important to have high energy so that  
the results in the current fragmentation region 
can be considered as purely from the struck quark 
(anti-quark) fragmentation. 
In this case, there is no new 
result for anti-neutrino charged current interactions
compared with those for the corresponding neutrino reactions. 
The differences come from the valence-quark contributions 
which are small at very high energies where small $x$ dominate.
In view of this and the difficulties in performing 
such experiments in the near future, we 
will not discuss this reaction in the next section.

\subsection{Models for $\Delta D_f^H(z)$}

There exist many different approaches for $\Delta D_f^H(z)$ 
in literature \cite{GH93,BL98,Kot98,Florian98,LL00,LXL01,XLL02,LL02,
Ma:1999wp,Ma:1998pd,BQMa01,Ma:2000cg,Ma:2000uu}.
We summarize the key points of some of them in the following.

\subsubsection{Calculation of $\Delta D_f^H(z)$ 
according to the origin of $H$}

In \cite{GH93,BL98,LL00,LXL01,XLL02,LL02}, 
$\Delta D_f^H(z)$ has been calculated according to the origins of $H$. 
The produced $H$'s are divided into the following four categories:
(A) those are directly produced and contain $q_f$; 
(B) decay products of polarized heavy hyperons; 
(C) those are directly produced and do not contain $q_f$; 
(D) decay products of unpolarized heavy hyperons. 
This is to say that we divide further
\begin{equation}
D_f^H(z;\text{dir})=D_f^{H(A)}(z)+D_f^{H(C)}(z);
\end{equation}
\begin{equation}
D_f^H(z;\text{dec})=D_f^{H(B)}(z)+D_f^{H(D)}(z).
\end{equation}
For the polarized case, 
\begin{equation}
\Delta D_f^H(z;\text{dir})=\Delta D_f^{H(A)}(z)+\Delta D_f^{H(C)}(z);
\end{equation}
\begin{equation}
\Delta D_f^H(z;\text{dec})=\Delta D_f^{H(B)}(z)+\Delta D_f^{H(D)}(z).
\end{equation}

It is assumed that, 
\begin{equation}
\Delta D_f^{H(A)}(z)=t^F_{H,f} D_f^{H(A)}(z),
\end{equation}
\begin{equation}
\Delta D_f^{H(C)}(z)=\Delta D_f^{H(D)}(z)=0.
\end{equation}
Here, $t^F_{H,f}$ is a constant and is taken as, 
\begin{equation}
t_{H,f}^F=\Delta Q_f/n_f
\end{equation} 
where $\Delta Q_f$ and $n_f$ are the fractional contribution of spin of 
quark with flavor $f$ to the spin of $H$ and the number of valence-quarks 
of flavor $f$ in $H$. Clearly, $Q_f$ is different in the SU(6) picture 
from those drawn from polarized deeply inelastic scattering (DIS) data. 
They are given in e.g. Table I of \cite{LL00}.

The decay contribution part 
$\Delta D_f^{H(B)}(z)$ is calculated using Eq.(5), i.e., 
\begin{equation}
\Delta D_f^{H(B)}(z)
=\sum_j\int dz' t^{D}_{H,H_j}
K_{H,H_j}(z,z')[\Delta D_f^{H_j(A)}(z')+\Delta D_f^{H_j(B)}(z')],
\end{equation}

We should note that, in the Feynman-Field type of 
recursive cascade fragmentation model, 
$D_f^{H(A)}(z)$ is nothing else but the 
probability to produce a first rank $H$ with $z$.
It is usually denoted by $f_{q_f}^H(z)$ in such 
fragmentation models, i.e., 
\begin{equation}
D_f^{H(A)}(z)=f_{q_f}^H(z).
\end{equation}
It follows that,
\begin{equation}
D_f^{H(C)}(z)=D_f^H(z;\text{dir})-f_{q_f}^H(z).
\end{equation}
In this case, we have,
\begin{equation}
\Delta D_f^{H(A)}(z)=t_{H,f}^Ff_{q_f}^H(z),
\end{equation}
\begin{equation}
\Delta D_f^{H(B)}(z)=\sum_j\int dz' t^{D}_{H,H_j}
K_{H,H_j}(z,z')[t_{H,f}^Ff_{q_f}^H(z)+\Delta D_f^{H_j(B)}(z')].
\end{equation}

In this model, for $\bar q_f\to HX$, we should have, 
\begin{equation}
D_{\bar f}^{H(A)}(z)=0,
\end{equation}
\begin{equation}
\Delta D_{\bar f}^H(z)=0.
\end{equation}

We recall that $f_{q_f}^H(z)$ is a quite essential input in 
such recursive cascade models. 
In most cases, there exist explicit expression for it.  
Hence, the $z$ dependence of $\Delta D$ in this model is determined by 
the fragmentation model for unpolarized case, which are empirically known
from the experimental facts for unpolarized reactions.
The only unknown is the spin transfer constant $t^F_{H,f}$ for 
type (A) hyperon which influences mainly the magnitudes of the 
polarizations of the hyperons. 
So the ingredients of this model can be tested separately by 
testing the $z$ dependence 
and magnitudes of the polarizations.

In \cite{GH93,BL98,LL00,LXL01,XLL02,LL02}, 
calculations have been carried out using a Monte Carlo event 
generator \cite{PYTHIA}
{\sc jetset} for $e^+e^-$ annihilation and {\sc pythia} for $lp$ or $pp$ collisions 
based on Lund fragmentation model \cite{lund}.
A set of formulas are given there which are more convenient for
Monte Carlo calculations. 

\subsubsection{Calculation of $\Delta D_f^H(z)$ using Gribov relation}

In \cite{Ma:1999wp,Ma:1998pd,BQMa01,Ma:2000cg,Ma:2000uu}, 
$\Delta D_f^H(z)$ has been calculated using 
the following proportionality relation, i.e., 
$D_f^H(z)\propto q_f^H(z)$ and 
$\Delta D_f^H(z)\propto \Delta q_f^H(z)$,
which they referred as ``Gribov relation'' since it was 
first shown in \cite{Gribov} by Gribov and collaborator in 1971.
In terms of the language given above, this relation, 
if true, should be valid for the directly produced part, 
i.e. 
\begin{equation}
D_f^H(z;\text{dir})\propto q_f^H(z),
\end{equation}
\begin{equation}
\Delta D_f^H(z;\text{dir})\propto \Delta q_f^H(z).
\end{equation}
The decay contribution parts can be calculated using Eq.(5). 

For $\bar q_f\to HX$, we have similar results, i.e. 
\begin{equation}
D_{\bar f}^H(z;\text{dir})\propto \bar q_f^H(z),
\end{equation}
\begin{equation}
\Delta D_{\bar f}^H(z;\text{dir})\propto \Delta \bar q_f^H(z).
\end{equation}

In \cite{Ma:1999wp,Ma:1998pd,BQMa01,Ma:2000cg,Ma:2000uu}, 
as an approximation, decay contributions are not considered. 

\subsubsection{Other models}

Other models for $\Delta D_f^H(z)$ have 
also discussed in the literature \cite{Kot98,Florian98}.
They are essentially different combinations of the two models 
discussed above. 
We will not go to the details of these models but refer 
the interested readers to the references.

\section{Results and discussions}

By applying the formulas presented in the last section to different reactions, 
we obtain the results for the polarization of hyperons and 
antihyperons in these reactions. 
We use the model for $\Delta D_f^H(z)$ as described 
in the first subsection of Sec. II C, i.e. to calculate $\Delta D_f^H(z)$ 
according to the origin of $H$.
The different contributions to $D_f^H(z)$ 
are calculated using a Monte Carlo event generator 
{\sc{pythia}} \cite{PYTHIA} based on Lund string 
fragmentation model \cite{lund}.
The results for hyperon polarization are given 
in \cite{BL98,LL00,LXL01,XLL02,LL02}.
We present the results for antihyperons 
and compare them with those for the corresponding hyperon 
in the following.  
As we have shown in Section IIB, 
the results for antihyperons in 
$e^+e^-$ annihilations differ from those 
for the corresponding hyperons only by a minus sign. 
We will not repeat them here. 
We will present the results for 
$e^-+N\to e^-+\bar H+X$ and those for 
$\nu_\mu+N\to \mu^-+\bar H+X$, respectively, in the following. 

\subsection{$e^-+N\to e^-+\bar H+X$}

\begin{figure}
\includegraphics[scale=0.65]{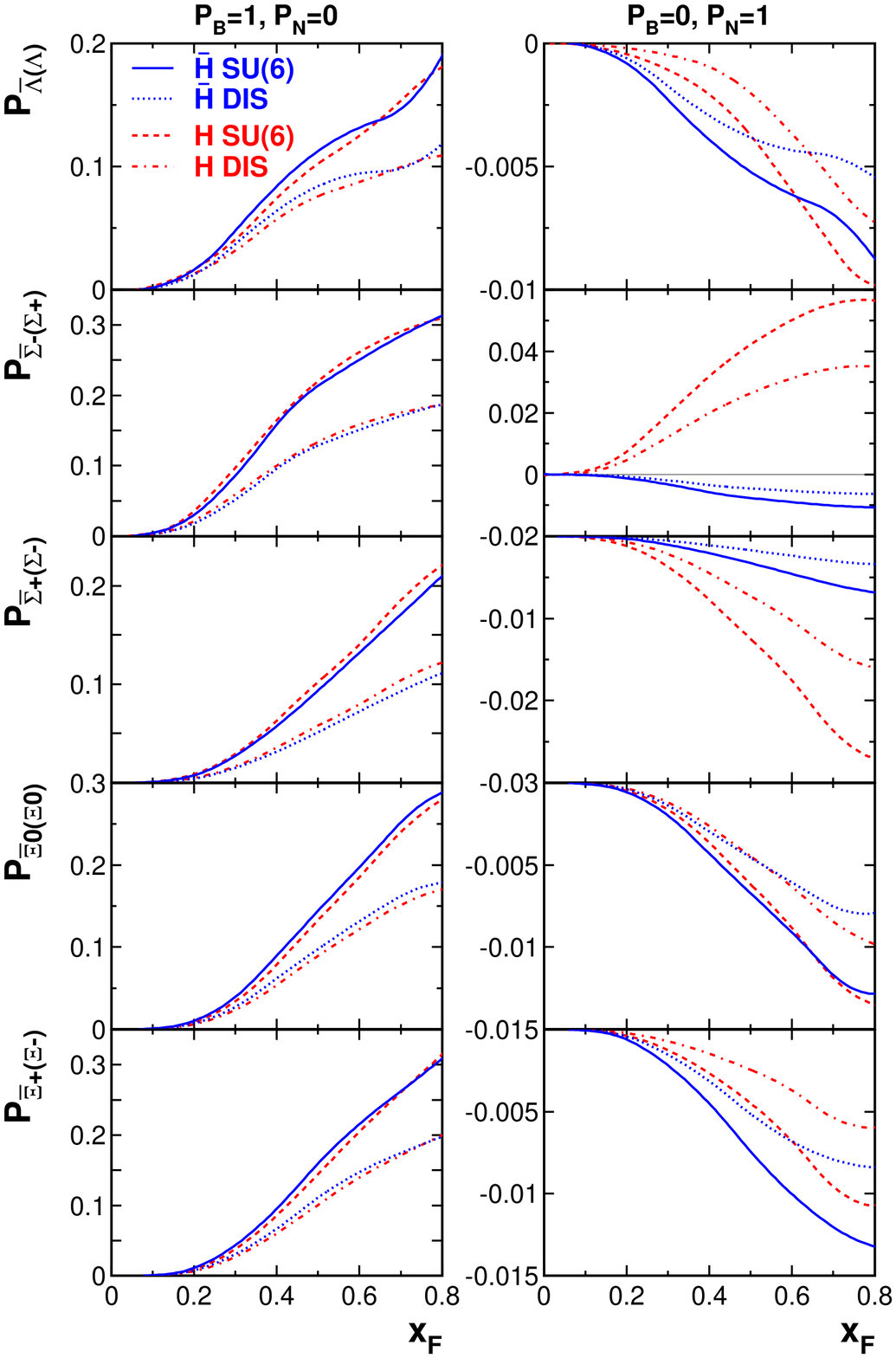}
\caption{\label{fig:muphigh}
Comparison of the 
polarizations of antihyperons 
with those of the corresponding hyperons 
as functions of $x_F$ 
in $e^- p\to e^-\bar H\text{(or\ }H\text{)} X$ 
at $E_{e^-}=500$ GeV. 
The solid and dashed lines denote the results obtained by using the SU(6) picture, 
while the dotted and dash-dotted lines denote those by using the DIS picture.  
The results for hyperons are the same as those presented in \cite{LXL01}. 
The quark distribution functions are taken from \cite{GRSV2000} and \cite{GRV98}. 
} 
\end{figure}

In Fig. \ref{fig:muphigh}, we show the results for antihyperons 
in $e^-+N\to e^-+\bar H+X$ with polarized beam or target.
Clearly the situation is completely the same for 
$e^++N\to e^++\bar H+X$ or $\mu^\pm+N\to \mu^\pm+\bar H+X$, 
where one virtual photon exchange plays the dominate role. 
The incident energy of electron is taken as $E=500$ GeV 
off a fixed target.
We take this electron energy to gurantee that 
the energy is already high enough 
that we need only take the struck quark or anti-quark 
fragmentation into account when we look at the 
current fragmentation region. 
The results are shown as functions of $z$, 
where $0.2<y<0.9 $, $Q^2>1 \text{GeV}^2$ and $W^2>1 \text{GeV}^2$. 
The results depend not much on the energy so long as 
$E$ is considerably large (say, larger than $200$GeV).

From Fig. \ref{fig:muphigh}, we see that there is indeed 
little difference between the results for hyperons and 
those for antihyperons at such energies for reactions 
with polarized beams. 
This confirms the qualitative expectations presented in Sec.IIB.
We also see that the polarizations are much larger in reactions 
with polarized beams than those for reactions where only the 
nucleon target is polarized.
The results in the latter case are very sensitive to the 
parametrization of polarized quark (anti-quark) distribution 
functions used. 
The relatively large difference between the results for 
hyperons and those for the corresponding antihyperons 
reflects the differences in the polarized 
quark (anti-quark) distributions.  

\begin{figure}
\includegraphics[scale=0.65]{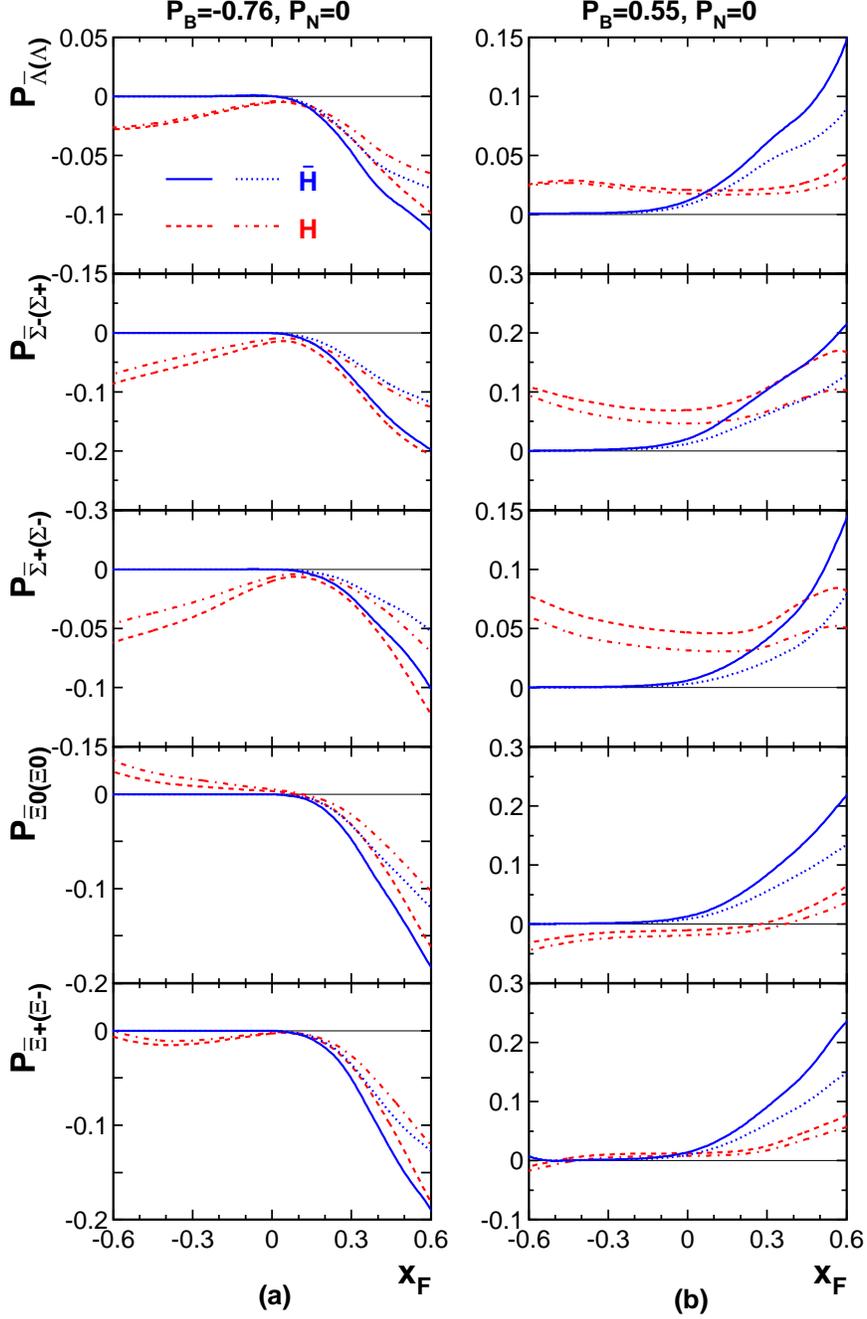}
\caption{\label{fig:mupcompasshermes}
Comparison of the 
polarizations of antihyperons 
with those of the corresponding hyperons 
as functions of $x_F$ 
in $\mu^+ p\to \mu^+\bar H\text{(or\ }H\text{)} X$ 
at $E_{\mu^+}=160$ GeV (in the left column) 
and that in $e^+ p\to e^+\bar H\text{(or\ }H\text{)} X$ 
at $E_{e^+}=27.6$ GeV (in the right column) 
when both the contributions from the fragmentation 
of the struck quark and that of the nucleon remnant are taken into account.  
The solid and dashed lines denote the results obtained by using the SU(6) picture, 
while the dotted and dash-dotted lines denote those by using the DIS picture.   }
\end{figure}

To compare the results with the results \cite{COMPASS}
from COMPASS at CERN, we lowered the energy 
to $E=160$GeV and obtained the results as shown 
in the left column of Fig. \ref{fig:mupcompasshermes}. 
Our results show that, at this energy, 
the difference between the results 
for $\Lambda$ and those for $\bar\Lambda$ 
polarization is also quite small, in particular, 
in the $x_F\sim 0$ region.  
The difference can be a little bit significant only for larger $x_F$. 
According to this result, we should not see a 
large difference between $P_\Lambda$ and $P_{\bar \Lambda}$ 
as indicated by the COMPASS data \cite{COMPASS}.

\begin{figure}
\includegraphics[scale=0.65]{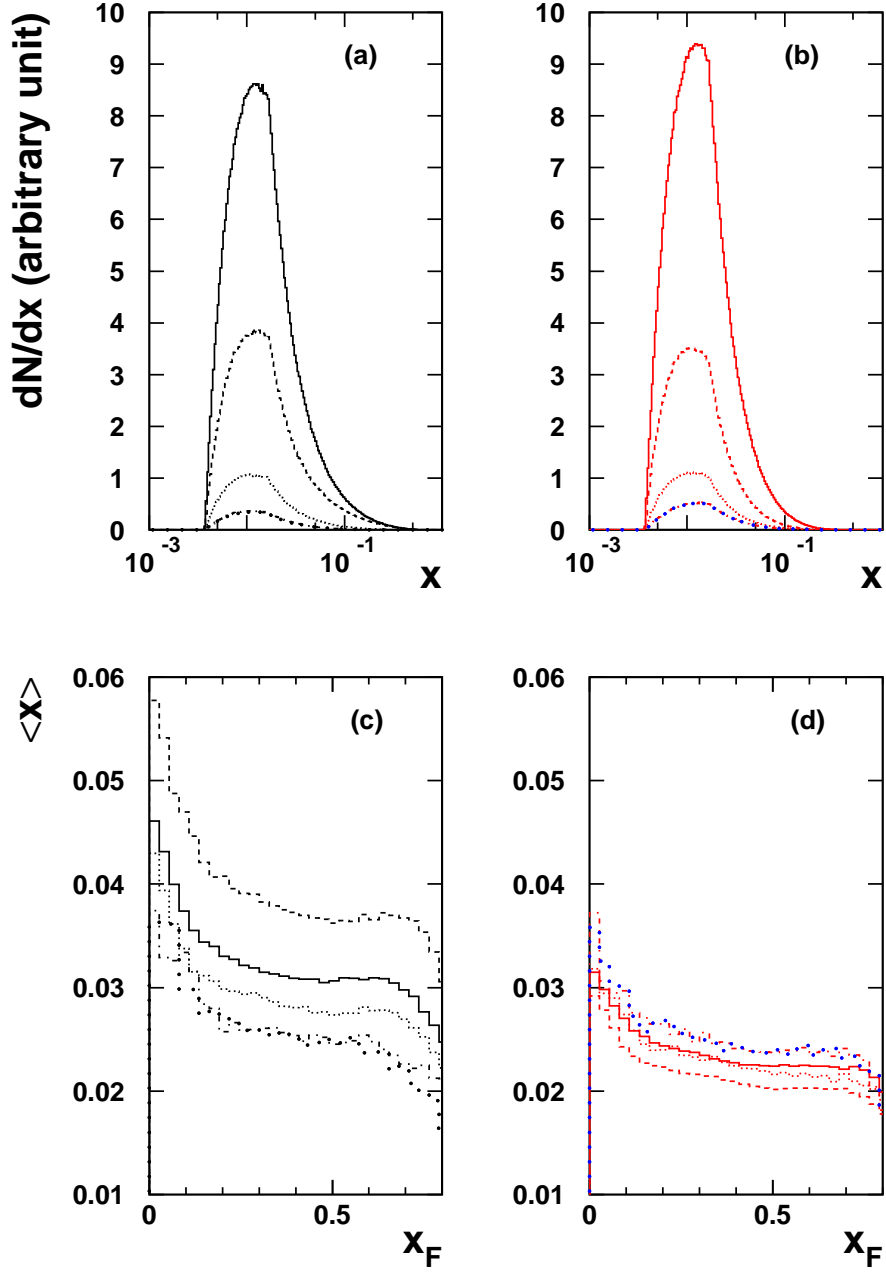}
\caption{\label{fig:averagexcompass}
The $x$-distribution obtained at the COMPASS energy 
for the struck quark or anti-quark that leads to 
the production of the hyperon [shown in (a)] 
and antihyperon [shown in (b)], and   
the average values of $x$ of such struck quarks or anti-quarks 
as functions of $x_F$ of the produced hyperons [in (c)] and 
antihyperons [in (d)] respectively. 
The solid, dashed, thick-dotted, dash-dotted and thin-dotted lines are 
for $\Lambda$, $\Sigma^+$, $\Sigma^-$, $\Xi^0$ and $\Xi^-$ or the 
corresponding antihyperons respectively 
(where the thin-dotted and dash-dotted lines are almost coincide with each other).  
}
\end{figure}

As we have discussed in last section, 
in the theoretical framework described in this paper, 
the following symmetries are supposed: 
(i) Charge conjugation symmetry in the fragmentation 
functions and decay processes [see Eqs.(14)--(17)]; and
(ii) charge conjugation symmetry in nucleon sea in particular 
$s(x)=\bar s(x)$ and $\Delta s(x)=\Delta \bar s(x)$.
In this case, 
the origins of the difference between the polarizations 
of hyperons and those of antihyperons can be only the 
contributions from the valence-quarks of the initial nucleons.
Our results show that the valence-quark contributions are 
already quite small at the COMPASS energy.
To see this explicitly, 
we have examined the $x$ values of the struck quark 
or antiquark for events with the kinematic constraints 
as imposed by COMPASS. 
(Here, $x$ is the Bjorken $x$ used in describing deeply 
inelastic lepton-nucleon scattering.) 
The results are shown in Fig.\ref{fig:averagexcompass}.
Here, in Fig.\ref{fig:averagexcompass}(a) and \ref{fig:averagexcompass}(b), 
we show the $x$-distribution of the struck 
quark and anti-quark that lead to the production of the 
hyperon and antihyperon, respectively; 
and in \ref{fig:averagexcompass}(c) and \ref{fig:averagexcompass}(d), 
we show the average values of $x$ of 
such struck quarks and anti-quarks as functions of $x_F$ of the 
produced hyperons or antihyperons.
We see that the $x$ value of the struck quark or anti-quark 
can be as small as $0.004$. 
We see also that $\langle x\rangle$ in this case is in general 
of the order of 0.01 for most of the $x_F$ values. 
In this $x$ region, $q(x,Q^2)$ is already dominated by 
the sea quark distribution $q_s(x,Q^2)$. 
There exists a valence-quark contribution $q_v(x,Q^2)$, 
but it is much smaller than $q_s(x,Q^2)$. 
This is why the obtained $P_{\bar H}$ differs 
little from $P_H$.

Another effect that relates to the valence-quark contribution 
and may cause a difference between $P_H$ 
and $P_{\bar H}$ in $e^-+N\to e^-+H({\rm or}\ \bar H)+X$ 
is the contribution from the hadronization of the 
remnant of target nucleon. 
It has been pointed out first in \cite{LL02} that 
contribution of the hadronization of target remnant 
is important to hyperon production 
even for reasonably large $x_F$ at lower energies. 
The effect has been confirmed by the calculations 
presented in \cite{Ellis2002}. 
It has been shown that \cite{LL02} at the CERN NOMAD energies, 
contribution from the hadronization of nucleon 
target remnant dominates hyperon production at $z$ around zero. 
It is impossible to separate the contribution of the 
struck quark fragmentation from those of the 
target remnant fragmentation.
Clearly, this effect can be different for 
hyperon and antihyperon production 
since the target remnant contribution comes mainly from 
the fragmentation of the valence diquark.
It contributes quite differently for hyperons and antihyperons.
To see whether this effect plays an important role at COMPASS energy, 
we have also calculated the contributions of target remnant fragmentation.
We found out that this contribution is already very small 
for $z>0$ at the COMPASS energy. 
To see the influence on hyperon polarizations, 
we have also included the contributions in the results shown in 
Fig. \ref{fig:mupcompasshermes}(a) by using a valence-quark model 
for the quark polarization in the remnant of the target 
as described in \cite{LL02}.
We see that there is indeed some influence at $x_F\sim 0$ 
for the polarization of the hyperons but the 
effect is already very small at the COMPASS energy.

After having made these checks, we are quite confident that 
the valence-quark contributions are quite small at the 
COMPASS energy. 
Hence, under the conjugation symmetries in fragmentation function, 
decay processes and in nucleon sea, there should not be a 
large difference between $P_\Lambda$ and $P_{\bar\Lambda}$ at COMPASS energy. 
A significant difference could be a signature for the violation 
of such conjugation symmetries. 
In view of the recent discussions on the asymmetric strange 
and antistrange sea \cite{BM96}, it would be very significant to 
check whether similar asymmetry is also possible for 
the polarized case.

We also made the calculations at HERMES energy, i.e. at $E=27.6$GeV.
The results are shown in the right column of 
Fig. \ref{fig:mupcompasshermes}. 
We found out that at this energy, the major contribution comes 
mainly from the $x$-region of $0.02<x<0.8$. 
In this $x$ region, valence-quark plays an important role. 
Hence, we obtained a significant difference 
between the polarizations of antihyperons and those of hyperons 
in the current fragmentation region. 
We see, in particular, that the difference for 
$\Lambda$ and $\bar\Lambda$ is indeed as expected 
in the qualitative analysis given in Sec. II B 2.

\begin{figure}
\includegraphics[scale=0.65]{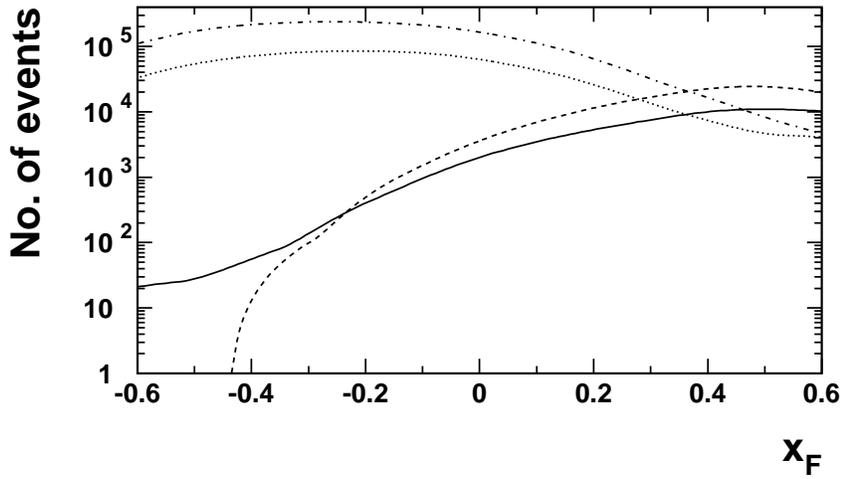}
\caption{\label{fig:muphermesdistribution}
Different contributions to $\Lambda$ in $e^+ p\to e^+\Lambda X$ 
at $E_{e^+}=27.6$ GeV. 
Here, the solid, dotted, dashed and dash-dotted lines denote respectively 
(1) directly produced and contain the struck quark; 
(2) directly produced and contain a $u$(or $d$)-quark 
in $\text{(}uu\text{)}_1$(or $\text{(}ud\text{)}_1$); 
(3) decay products of hyperons which contain the struck quark; 
(4) decay products of hyperons which contain the $\text{(}uu\text{)}_1$(or $\text{(}ud\text{)}_1$) 
or a $u$(or $d$) in the $\text{(}uu\text{)}_1$(or $\text{(}ud\text{)}_1$).  }
\end{figure}

We also calculated the contribution from target remnant 
fragmentation and found that it is important at HERMES energy. 
This can be seen clearly in 
Fig. \ref{fig:muphermesdistribution}, 
where different contributions to $\Lambda$ at this energy are shown. 
We see clearly that the contribution from target remnant fragmentation 
is much higher than that from the struck quarks 
in the region near $x_F=0$. 
To show the influence of this effect on the polarizations, 
we calculate these contributions from target remnant fragmentation 
using a valence-quark model for the polarizations of the quarks 
in the remnant of the nucleon as described in \cite{LL02}. 
The results are added to the struck quark fragmentation contribution 
and are shown in Fig. \ref{fig:mupcompasshermes}(b). 
We see that the contribution is very important in the $x_F\sim 0$ region.
It is therefore meaningless to factorize the $\Lambda$ production 
cross section as one usually does at very high energies.   
On the other hand, the influence from target remnant fragmentation 
to antihyperon polarization is negligibly small.
This is another reason for the differences between 
the polarizations of hyperons and those of antihyperons 
shown in Fig. \ref{fig:mupcompasshermes}(b) in particular at the $x_F\sim 0$ region. 
 
\subsection{$\nu_\mu+N\to \mu^-+\bar H+X$} 

\begin{figure}
\includegraphics[scale=0.65]{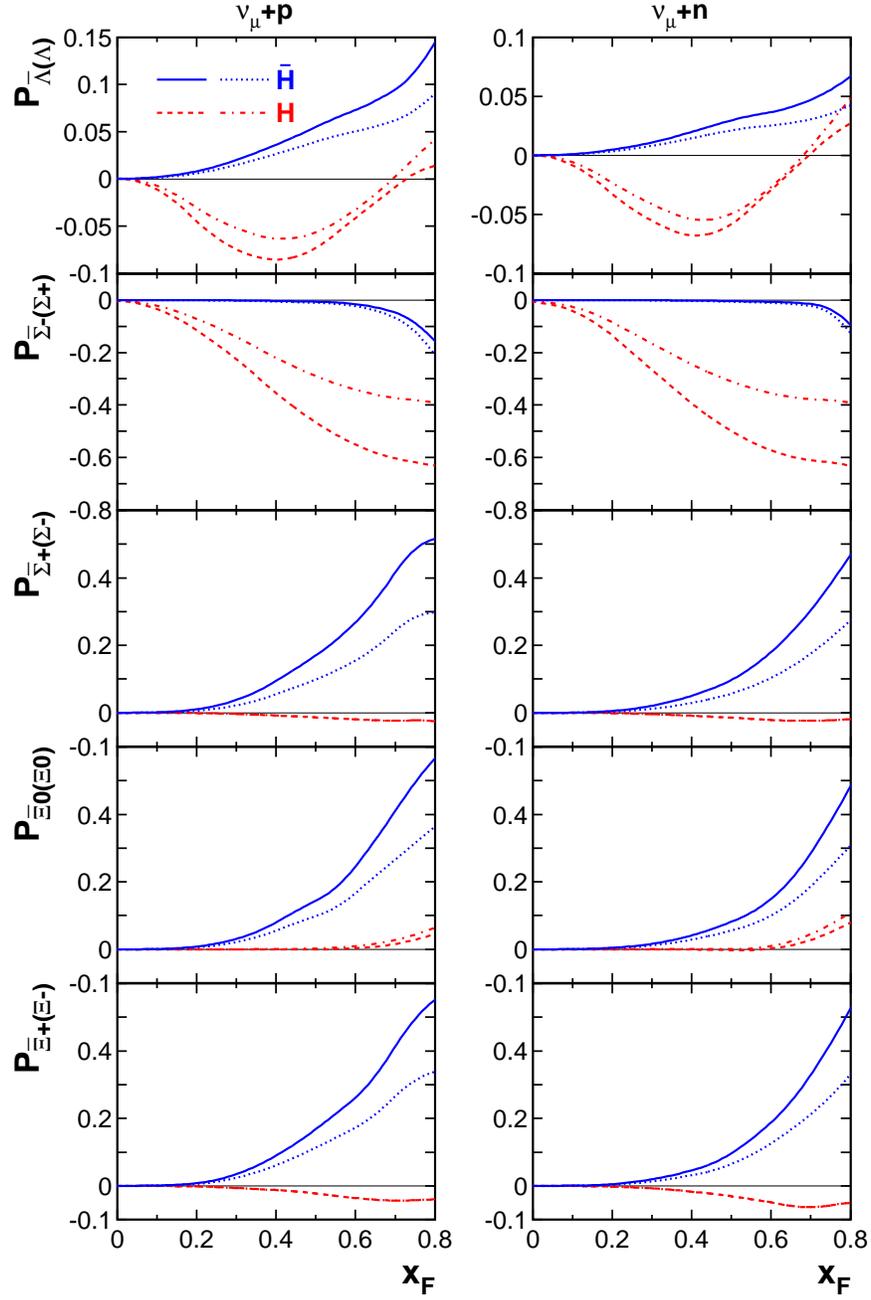}
\caption{\label{fig:neupnhigh}
Comparison of the 
polarization of antihyperons 
with those of the corresponding hyperons 
as functions of $x_F$ 
in $\nu_\mu N\to \mu^-\bar H\text{(or\ }H\text{)} X$ 
at $E_\nu=500$ GeV. 
The solid and dashed lines denote the results obtained by using the SU(6) picture, 
while the dotted and dash-dotted lines denote those by using the DIS picture.  }
\end{figure}

In Fig. \ref{fig:neupnhigh}, we show the results obtained 
for $\nu_\mu+N\to \mu^-+\bar{H}\text{(or }H\text{)}+X$ 
at an incident muon-neutrino beam energy $E=500\ \text{GeV}$ 
off a fixed target. 
We calculated for proton and neutron target and different hyperons 
and antihyperons.  
To obtain the results for $\Lambda$, 
we simply take $t^D_{\Lambda,\Lambda_c}=1$ 
as we did in [7].
From these figures, we see that the results for hyperons 
and those for antihyperons are indeed significantly different from each other. 
These features are quite different from that in electron nucleon collision 
where hyperon and antihyperon polarizations are essentially the same at 
very high energies.  
The qualitative features are the same as we obtained 
in the qualitative analysis using Eqs.(65) and (66) in the last section.

\begin{figure}
\includegraphics[scale=0.65]{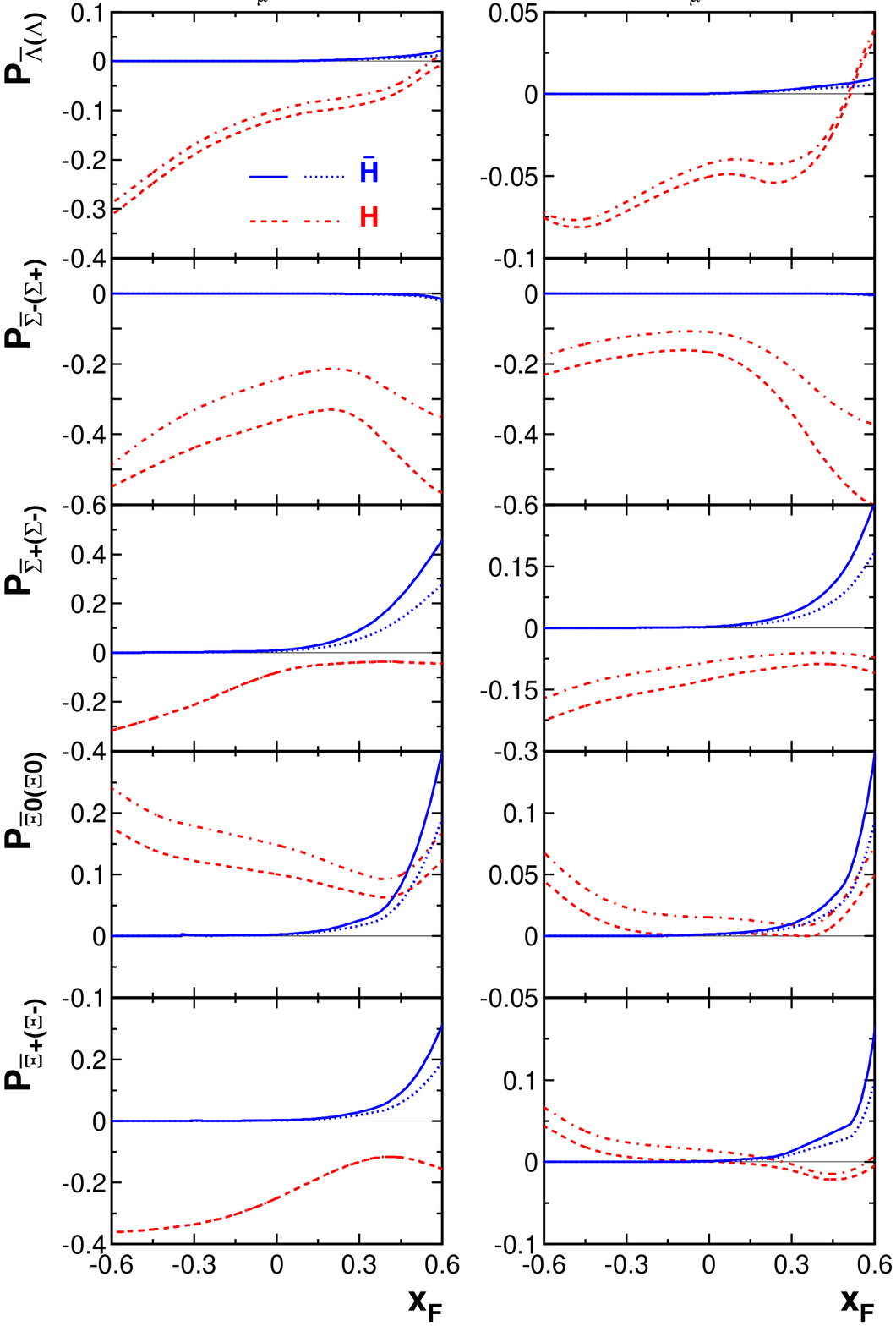}
\caption{\label{fig:neupnnomad}
Polarizations of antihyperons 
compared with those of the corresponding hyperons 
as functions of $x_F$ 
in $\nu_\mu N\to \mu^-\bar H\text{(or\ }H\text{)} X$ 
at $E_\nu=44$ GeV when both the contributions from the fragmentation 
of the struck quark and that of the nucleon remnant are taken into account.  
The solid and dashed lines denote the results obtained by using the SU(6) picture, 
while the dotted and dash-dotted lines denote those by using the DIS picture. }
\end{figure}

We also calculated the polarizations for antihyperons 
at the CERN NOMAD energies. 
The results are shown in Fig. \ref{fig:neupnnomad}. 
As it has been pointed out in \cite{LL02}, at this energy, 
contribution from the fragmentation of the remnant of target 
nucleon to hyperon production is very important.
We also included this contribution from a rough estimation 
by using a valence-quark model for the polarization of the 
quarks in the target remnant as in \cite{LL02}. 
For this reason, the polarizations obtained for hyperons 
at this energy differ significantly from those obtained 
at very high energis as shown in Fig. \ref{fig:neupnhigh}.
But for antihyperons, this contribution is small 
thus the difference between the results and those 
shown in Fig. \ref{fig:neupnhigh} is small.  
This is also one of the reasons for the differences 
between the results for hyperons and those for antihyperons, 
in particular, in the $x_F\sim 0$ region. 
The qualitative features are consistent with the NOMAD data \cite{NOMAD00} 
and it will be interesting to make high statistics check in the future. 

\section{Summary and outlook}

In summary, we have calculated the polarizations of antihyperons 
in lepton induced reactions, in particular $l+p\to l^{'}+\bar H+X$ 
at different energies with polarized beams,  
using different models for spin transfer for fragmentation. 
The results show little difference between the polarization of 
antihyperon and that of the corresponding hyperon in 
$e^-+N\to e^-+\bar H (\rm{or}\ H)+X$ at COMPASS or higher energies 
when the charge conjugation symmetries for fragmentation, decay 
and nucleon sea are assumed.
But there are in general large differences between those for 
antihyperons and the corresponding hyperons in neutrino 
induced charged current reactions since the 
flavors of dominant fragmenting quarks and those of the anti-quarks 
and their polarizations are different.
A detailed discussion is given and the results 
can be tested by future experiments.

\section*{Acknowledgments}

We thank M.G. Sapozhnikov of the COMPASS Collaboration for suggesting 
that we make such calculations and for stimulating discussions.
This work was supported in part by
the National Science Foundation of China (NSFC) with grant Nos. 10135030 and 
10405016.

\newpage

\begin{thebibliography}{99}
\bibitem{ft} The spin transfer factor $t^D_{H_i,H_j}$ can in general be dependent of 
 the momentum of $H_i$ and $H_j$ for parity violating decay. We, as usually do in 
 literature, simply neglect this effect in this paper and 
 leave the topic for further separate study.
\end{thebibliography}
\end{document}